\documentclass[12pt,epsf]{article}
\usepackage{graphicx}
\usepackage{amsmath,amssymb}
\begin{document}

\def\a{\alpha}
\def\b{\beta}
\def\c{\varepsilon}
\def\d{\delta}
\def\e{\epsilon}
\def\f{\phi}
\def\g{\gamma}
\def\h{\theta}
\def\k{\kappa}
\def\l{\lambda}
\def\m{\mu}
\def\n{\nu}
\def\p{\psi}
\def\q{\partial}
\def\r{\rho}
\def\s{\sigma}
\def\t{\tau}
\def\u{\upsilon}
\def\v{\varphi}
\def\w{\omega}
\def\x{\xi}
\def\y{\eta}
\def\z{\zeta}
\def\D{\Delta}
\def\G{\Gamma}
\def\H{\Theta}
\def\L{\Lambda}
\def\F{\Phi}
\def\P{\Psi}
\def\S{\Sigma}

\def\o{\over}
\def\beq{\begin{eqnarray}}
\def\eeq{\end{eqnarray}}
\newcommand{\gsim}{ \mathop{}_{\textstyle \sim}^{\textstyle >} }
\newcommand{\lsim}{ \mathop{}_{\textstyle \sim}^{\textstyle <} }
\newcommand{\vev}[1]{ \left\langle {#1} \right\rangle }
\newcommand{\bra}[1]{ \langle {#1} | }
\newcommand{\ket}[1]{ | {#1} \rangle }
\newcommand{\EV}{ {\rm eV} }
\newcommand{\KEV}{ {\rm keV} }
\newcommand{\MEV}{ {\rm MeV} }
\newcommand{\GEV}{ {\rm GeV} }
\newcommand{\TEV}{ {\rm TeV} }
\def\diag{\mathop{\rm diag}\nolimits}
\def\Spin{\mathop{\rm Spin}}
\def\SO{\mathop{\rm SO}}
\def\O{\mathop{\rm O}}
\def\SU{\mathop{\rm SU}}
\def\U{\mathop{\rm U}}
\def\Sp{\mathop{\rm Sp}}
\def\SL{\mathop{\rm SL}}
\def\tr{\mathop{\rm tr}}

\def\IJMP{Int.~J.~Mod.~Phys. }
\def\MPL{Mod.~Phys.~Lett. }
\def\NP{Nucl.~Phys. }
\def\PL{Phys.~Lett. }
\def\PR{Phys.~Rev. }
\def\PRL{Phys.~Rev.~Lett. }
\def\PTP{Prog.~Theor.~Phys. }
\def\ZP{Z.~Phys. }


\baselineskip 0.7cm

\begin{titlepage}

\begin{flushright}
IPMU 15-0043\\
\end{flushright}

\vskip 1.35cm
\begin{center}
{\large \bf
Seminatural SUSY from $E_7$ Nonlinear Sigma Model
}
\vskip 1.2cm
  Keisuke Harigaya$^{a,b}$,  Tsutomu T. Yanagida$^{b}$ and Norimi Yokozaki$^{c}$
\vskip 0.4cm

{\it
$^a$ ICRR, the University of Tokyo, Kashiwa, 277-8582, Japan \\
$^b$ Kavli IPMU (WPI), UTIAS, the University of Tokyo, Kashiwa, 277-8583, Japan \\
$^c$ INFN, Sezione di Roma, Piazzale A. Moro 2, I-00185 Roma, Italy
}

\vskip 1.5cm

\abstract{
We present a new focus point
supersymemtry breaking scenario based on the supersymmetric $E_7$ non-linear sigma model.
In this non-linear sigma model, squarks and sleptons are identified with (pseudo) Nambu-Goldstone bosons.
Their masses are generated only radiatively
through gauge and yukawa interactions,
and they are much smaller than the gravitino and gaugino masses at a high energy scale.
On the other hand, Higgs doublets belong to matter multiplets and hence may have unsuppressed supersymmetry-breaking soft masses.
We consider their masses to be equal to the gravitino mass at the high energy scale,
assuming the minimal Kahler potential for Higgs doublets.
We show that the fine-tuning measure of the electroweak symmetry breaking scale is reduced significantly to $\Delta=30\mathchar`-70$,
if the ratio of the gravitino mass to the gaugino mass is around $5/4$.
Also, the prospects of the discovery/exclusion of supersymmetric particles at the Large Hadron Collider and dark matter direct detection experiments are discussed.
}

\end{center}

\vskip 1.5cm

\begin{flushright}
{\em Prepared for submission to Progress of Theoretical and Experimental Physics}
\end{flushright}

\end{titlepage}

\setcounter{page}{2}

\section{Introduction}

The supersymmetric (SUSY) $E_7$ non-linear sigma (NLS) model based on $E_7/SU(5)\times U(1)^3$~\cite{Kugo:1983ai,Yanagida:1985jc} is attractive
since it accomodates three generations of quarks and leptons as Nambu-Goldstone (NG) chiral multiplets~\cite{Buchmuller:1982xn,Buchmuller:1982tf}.
The NLS model approach based on exceptional groups has a potential for predicting the maximal number of generations because the maximal volume of exceptional groups is limited by $E_8$.
In fact, we have four generations and one anti-generation  in  $E_8$ NLS models.
Thus, the net generation number is also three.
Futhermore, the NLS model may explain the observed small yukawa coupling constants for the first, second and third generations because of the celebrated low energy theorem \cite{Adler:1964um}. It may be intriguing that the basic structure of the $E_7$ NLS model does not change much even if we replace the $E_7$ by $E_{7(7)}$ symmetry found in the $N=8$ supergravity~\cite{Cremmer:1979up}.

We identify the unbroken subgroup $SU(5)$ with the gauge group of grand unification (GUT) and assume that the $E_7$ is an exact global symmetry in the limit where all yukawa and gauge coupling constants vanish.
We consider that all SUSY-breaking soft masses for squarks and sleptons are suppressed at some high energy scale such as the GUT scale.
On the other hand, gauginos obtain SUSY breaking masses $M_{1/2}$ of order of the gravitino mass, as in usual gravity mediation.
Then squarks and sleptons obtain their soft masses mainly from radiative corrections by gaugino loops, which is nothing but so-called gaugino mediation~\cite{Inoue:1991rk,Kaplan:1999ac,Chacko:1999mi}.
It is remarkable that gaugino mediation models are free from the serious flavor-changing neutral current problem, since the radiatively induced soft masses of squarks and sleptons are generation-independent.

The above $E_7$ NLS model also has one NG chiral multiplet ${\bf 5}'$ beside three generations of quarks and leptons.
Gauge and NLS model anomaly cancellation require an addtional matter multiplet ${\bf \bar{5}}'$~\cite{Yanagida:1985jc}.
It is natural that the NG multiplet ${\bf 5}'$ acquires an invariant mass together with  ${\bf \bar{5}}'$.
Therefore, massless NG multiplets are only three generations of quarks and leptons%
\footnote{The Kahler manifold  $E_7/SU(5)\times U(1)^3$ also accomodates three right-handed neutrinos as NG chiral multiplets \cite{Sato:1997hv,Evans:2013uza}.
If they have Majorana masses at an intermediate scale, mass parameters of right-handed neutrinos and ${\bf 5}' {\bf \bar{5}}'$ are regarded as explicit breaking parameters of the $E_7$ symmetry.}.

In addition to the NG chiral multiplets, 
we introduce a pair of Higgs multiplets, $H_u$ and $H_d$.
Since they are not NG chiral multiplets, SUSY breaking soft masses of them are not suppressed at the high energy scale.
We assume that their masses are given by the gravitino mass $m_{3/2}$, taking the minimal Kahler potential for them.
Obviously, those soft masses do not disturb the flavor-independent nature of
the soft masses of squarks and sleptons in the first and the second generations,
since their yukawa couplings are very small.

The purpose of this paper is to show the presence of a focus point~\cite{Feng:1999mn} when the mass ratio $r=m_{3/2}/M_{1/2} \simeq 6/5\mathchar`-4/3$.
This is very much similar to the focus point in gaugino mediation~\cite{Yanagida:2013ah}, where a non-universal gaugino mass spectrum is however required.\footnote{
It has been shown that the non-universal gaugino masses relax the fine-tuning of the electroweak symmetry breaking in general gravity mediation~\cite{nonuniv}. 
}
We find  that the required degree of fine-tuning is indeed quite mild as a few $\%$ (so-called $\Delta =50\mathchar`-100$).
We also discuss the potential of  the Large Hadron Collider (LHC) for testing the present model.

\section{$E_7/SU(5)\times U(1)^3$ NLS model in supergravity}

In this section, we review an $E_7/SU(5)\times U(1)^3$ NLS model in supergravity.
We first show that the $E_7/SU(5)\times U(1)^3$ NLS model accommodates three generations of quarks and leptons.
Then we discuss the mass spectrum of
minimal SUSY standard model (MSSM) particles at the tree-level.

\subsection{Three generations as NG chiral multiplets}

Let us construct the Lie algebra of $E_7$ by considering a maximal subgroup $SU(8)$.
Generators of $E_7$ are decomposed into $63$ generators of $SU(8)$, ${\hat{T}_I}^J$, and $70$ anti-symmetric tensors of $SU(8)$, $E_{IJKL}$ ($I,J,K,L = 1\mathchar`- 8$).
The anti-symmetric tensors obey a reality constraint, ${E_{IJKL}}^* = \epsilon^{IJKLMNOP}E_{MNOP}/ 4!$.
They satisfy the following algebra;
\begin{eqnarray}
\label{eq:E7 algebra}
\left[ {\hat{T}_I}^J, {\hat{T}_K}^L \right] &=& {\delta_K}^J {\hat{T}_I}^L - {\delta_I}^L {\hat{T}_K}^J,  \\
\left[ {\hat{T}_I}^J, E_{KLMN}\right] 
&=& {\delta_K}^J E_{ILMN} + {\delta_L}^J E_{KIMN} +{\delta_M}^J E_{KLIN} + {\delta_N}^J E_{KLMI}
- \frac{1}{2} {\delta_I}^J E_{KLMN},
\nonumber \\
\left[ E_{IJKL}, E_{MNOP}  \right] &= &
\frac{1}{2} \left( {\hat{T}_I}^Q \epsilon_{QJKLMNOP} + {\hat{T}_J}^Q \epsilon_{IQKLMNOP}  + {\hat{T}_K}^Q \epsilon_{IJQLMNOP}  + {\hat{T}_L}^Q \epsilon_{IJKQMNOP} \right) \nonumber\\
&& - \frac{1}{2} \left( {\hat{T}_M}^Q \epsilon_{IJKLQNOP} + {\hat{T}_N}^Q \epsilon_{IJKLMQOP} +{\hat{T}_O}^Q \epsilon_{IJKLMNQP} +{\hat{T}_P}^Q \epsilon_{IJKLMNOQ}   \right). \nonumber 
\end{eqnarray}

For clarity, we first consider an $E_7/SU(5)\times SU(3) \times U(1)$ NLS model~\cite{Kugo:1983ai}.
$133-24-8-1 = 100$ broken generators are labelled by $SU(5)$ indices $a,b,c,\cdots (= 1\mathchar`-5)$ and $SU(3)$ indices $i,j,k,\cdots (=1,2,3)$ as
\begin{eqnarray}
{\hat{T}_a}^i &\equiv& {X_a}^i,~ {\hat{T}_i}^a \equiv {\bar{X}_i}^a,\\
-\frac{1}{4!}\epsilon^{abcde}E_{bcde} &\equiv& X^a,~
- \frac{1}{3!} \epsilon^{abcde}E_{cdei} \equiv X_i^{ab},~
\frac{1}{2!} \epsilon^{ijk}E_{abjk} \equiv \bar{X}_{ab}^i,~
\frac{1}{3!} \epsilon^{ijk}E_{aijk} \equiv \bar{X}_a. \nonumber
\end{eqnarray}
Unbroken generators of $SU(5)\times SU(3) \times U(1)$ are given by
\begin{eqnarray}
{T_a}^b \equiv {\hat{T}_a}^b-\frac{1}{2} \sqrt{\frac{3}{10}}T,~~
{T_i}^i \equiv {\hat{T}_i}^j-\frac{1}{2} \sqrt{\frac{5}{6}}T,~~
T \equiv 2 \sqrt{\frac{2}{15}} {\hat{T}_a}^a.
\end{eqnarray}

$E_7/SU(5)\times SU(3) \times U(1)$ Kahler manifold is parameterized by complex parameters $(\phi_a^i, \phi_i ^{ab}, \phi^a)$ associated with broken generators $(\bar{X}_i^a, \bar{X}_{ab}^i, \bar{X}_a)$~\cite{Kugo:1983ai,Itoh:1985ha}.
$(\phi_a^i, \phi_i ^{ab}, \phi^a)$ transform under $SU(5)\times SU(3) \times U(1)$ as
\begin{eqnarray}
\phi_a^i~:~({\bf \bar{5}},{\bf 3},2),~~ \phi_i^{ab}~:~({\bf 10},{\bf \bar{3}},1),~~\phi^a~:~ ({\bf 5},{\bf 1},3).
\end{eqnarray}
It should be noted that 3 copies of ${\bf\bar{5}}$ and ${\bf 10}$ arise as NG fields.
We identify them with 3 generations of quark ($Q,\bar{u},\bar{d}$) and lepton ($L,\bar{e}$) chiral fields,
by gauging $SU(5)$.
Note that $E_7$ symmetry is explicitly broken by gauge couplings.

An $E_7/SU(5)\times U(1)^3$ NLS model~\cite{Yanagida:1985jc,Sato:1997hv} is obtained by breaking $SU(3)$ down to $U(1)^2$.
Three NG chiral fields associated with  $8-2 = 6$ broken generators of $SU(3)$ are identified with $3$ generations of right-handed neutrinos ($N$).

In addition to the NG chiral fields mentioned above,
we need an additional ${\bf \bar{5}}'$ to cancel the $SU(5)$ gauge the NLS anomalies~\cite{Yanagida:1985jc}.
One may identify doublets in $\phi^a$ and ${\bf \bar{5}}'$ as Higgs fields $H_u$ and $H_d$.
In this case, scalar soft mass squared of MSSM chiral multiplets except for that of $H_d$ vanish at the tree-level (see Sec.~\ref{sec:soft mass}).
Then, with a specific relation between the wino and the gluino mass,
we obtain the focus point discovered in~\cite{Yanagida:2013ah}.

In this paper, instead, we assume that $\phi^a$ and ${\bf \bar{5}}'$ obtain their large Dirac mass term
and decouple from low energy dynamics.
We introduce Higgs doublets $H_u$ and $H_d$, in addition to the NG chiral fields and ${\bf \bar{5}}'$.
As we show in the next section, we have a focus point even in this case.

As usual, the yukawa coupling of $H_u$ and $H_d$ with quarks, leptons and right-handed neutrinos $N$ are given by
\begin{eqnarray}
W = y_u H_u Q \bar{u} + y_d H_d Q \bar{d} + y_e H_d L \bar{e} + y_N H_u L N,
\end{eqnarray}
where we have suppressed generation indices, for simplicity.
The yukawa couplings also break $E_7$ symmetry explicitly.

\subsection{Kahler potential of NG fields}

Here,
we explain properties of Kahler potentials of NLS models necessary for our discussion.
For the construction and the full expression of the Kahler potential, see~\cite{Itoh:1985ha}.

According to the general procedure presented in~\cite{Itoh:1985ha},
one can construct a real function ${\cal K}(\phi, \phi^\dag) = \phi^\dag \phi + \cdots $ of NG chiral fields $\phi$ whose transformation law under $E_7$ is given by
\begin{eqnarray}
\delta_X {\cal K}(\phi,\phi^\dag) =& f_X(\phi) + f_X(\phi)^\dag &: \text{broken symmetry},  \nonumber \\
\delta_T {\cal K}(\phi,\phi^\dag) =& 0 &: \text{unbroken symmetry}.
\end{eqnarray}
In global SUSY, ${\cal K}$ is identified with the Kahler potential because the holomorphic terms $f_X(\phi)$ do not contribute to the action.
In supergravity, however, the holomorphic terms do contribute to the action.
Thus we are led to introduce a chiral field $S$~\cite{Komargodski:2010rb,Kugo:2010fs}
whose transformation law under $E_7$ is defined by
\begin{eqnarray}
\delta_X S = - f_X(\phi).
\end{eqnarray}
Then the Kahler potential invariant under $E_7$ is given by
\begin{eqnarray}
\label{eq:E7 Kahler}
K(\phi,\phi^\dag, S,S^\dag) = F ({\cal K}(\phi,\phi^\dag) + S + S^\dag),
\end{eqnarray}
where $F(x) = x + \cdots $ is a real function of $x$.

\subsection{Soft masses of MSSM fields}
\label{sec:soft mass}

Let us derive soft masses of MSSM fields at the tree-level.
We solve the renormalization equation of soft masses in the next section,
regarding the tree-level soft masses as boundary conditions at a high-energy scale.

Due to the NG boson-nature, soft mass squared of squarks and sleptons vanish at the tree-level:
as we have discussed in the previous section, the Kahler potential of quarks and leptons are given by
\begin{eqnarray}
\label{eq:Kahler}
K_{q,l} = F(q^\dag q + S + S^\dag,\cdots) + (\text{higher order in } q),  
\end{eqnarray}
where $q$ denotes quarks and leptons collectively.
The ellipse denotes other fields e.g.~SUSY breaking fields.
Terms of higher order in $q$ are irrelevant for our discussion on soft masses and hence we ignore them.
From Eq.~(\ref{eq:Kahler}),
$A$ and $F$ terms of $q$ ($q,F^q$) and those of $S$ ($S, F^S$) enter the scalar potential in the following form;
\begin{eqnarray}
V = G (q q^\dag + S + S^\dag, F^q q^\dag + F^S, F^{q^\dag}q + F^{S^\dag}, F^q F^{q^\dag},\cdots),
\end{eqnarray}
where $G$ is some function and  the ellipse denotes dependence on other fields.
We have ignored the contribution from the superpotential of $q$, since it is irrelevant for soft mass squared.
Solving the equation of motion of $F^q$ and $F^s$, we obtain
\begin{eqnarray}
V = V (q q^\dag + S + S^\dag,\cdots),
\end{eqnarray}
where the ellipse denotes dependence on other fields.
The soft mass squared of $q$ is given by
\begin{eqnarray}
 m_q^2 = \frac{\partial}{\partial q} \frac{\partial}{\partial q^\dag} V|_{q=0} = \frac{\partial}{\partial S} V.
\end{eqnarray}
The right-handed-side vanishes at the vacuum of $S$.

We assume that the Kahler potential of $H_u$ and $H_d$ is  the minimal,
\begin{eqnarray}
K_h = H_u^\dag H_u + H_d^\dag H_d.
\end{eqnarray}
Then soft mass squared of $H_u$ and $H_d$ is given by the gravitino mass $m_{3/2}$;
\begin{eqnarray}
m_{H_u}^2 = m_{H_d}^2 = m_{3/2}^2.
\end{eqnarray}

Gaugino masses are given by couplings between gauge multiplets and the SUSY breaking field $Z$ in the gauge-kinetic function,
\begin{eqnarray}
\int {\rm d}^2 \theta (\frac{1}{g^2} + k Z ) W^\alpha W_\alpha,
\end{eqnarray}
where $g$ is the gauge coupling constant, $k$ is a constant, and $W^\alpha$ is the superfield strength of the gauge multiplets.
Assuming that MSSM gauge multiplets are unified to an $SU(5)$ gauge multiplet,
the universal gaugino masses are given by%
\footnote{
Universality of gaugino masses is not crucial for the focus point discussed in the next section (see Fig.~\ref{fig:delta_non_univ}).
}
\begin{eqnarray}
M_{1/2} = \frac{\sqrt{3}}{2} k g^2 K_{ZZ^\dag}^{-1/2} m_{3/2}
\end{eqnarray}
at the GUT scale, where $K_{ZZ^\dag}$ denotes the derivative of the Kahler potential with respect to $Z$ and $Z^\dag$.
Here, we assume that the SUSY is dominantly broken by the $F$ term of $Z$.

\section{Focus point for the electroweak symmetry breaking}

Let us first assume that the Kahler potential of
the SUSY breaking field $Z$ is the minimal one and its vacuum expectation value (VEV) is much smaller than the Planck scale.
In this special case,
trilinear A-terms almost vanish.
We discuss the case of non-vanishing A terms later.
As for the gaugino masses, we assume the universal gaugino mass $M_1=M_2=M_3= M_{1/2}$.
We also discuss the case of non-universal gaugino masses, where  we see the focus point behavior is maintained.
As we have shown in the previous section, soft masses of squarks and sleptons all vanish at the tree level. 
However, the global $E_7$ symmetry is not exact and hence it may be more natural to consider that they have non-vanishing small masses.
These non-vanishing soft masses are expected to be much smaller than the gravitino mass $m_{3/2}$~\cite{Evans:2013uza,Gaillard:2000fk,Harigaya:2015kfa}, and hence they have only small effects on the fine-tuning of the electroweak symmetry breaking (EWSB) scale (see Fig.~\ref{fig:m0}).
In this paper, we assume that squarks and sleptons have vanishing soft masses, for simplicity.
Thus, we have only three soft SUSY breaking masses,  $m_{3/2}(=m_{H_u}=m_{H_d})$, $M_{1/2}$ and $B_0 = B_\mu/\mu |_{M_{\rm inp}}$.%
\footnote{The Higgs $\mu$ and $B_\mu$ terms are assumed to arise from the Giudice-Masiero mechanism~\cite{giudice_masiero}. Then, $B_0$ is regarded as a free parameter.}
Here, $M_{\rm inp}$ is the mass scale where those soft SUSY breaking masses are set, and is taken as $M_{\rm inp}=10^{16}$ GeV.

The EWSB conditions are given by
\begin{eqnarray}
\frac{g_1^2 + g_2^2}{4} v^2 &=&  -\mu^2 
- \frac{(m_{H_u}^2  + \frac{1}{2 v_u}\frac{\partial \Delta V}{\partial v_u} ) \tan^2\beta}{\tan^2\beta-1} 
 + \, \frac{m_{H_d}^2 + \frac{1}{2 v_d}\frac{\partial \Delta V}{\partial v_d} }{\tan^2\beta-1}  \Bigr|_{M_{\rm IR}}, \nonumber \\
\frac{B \mu \,(\tan^2\beta+1)}{\tan\beta} &=&  m_{H_u}^2 +\frac{1}{2 v_u}\frac{\partial \Delta V}{\partial v_u} + m_{H_d}^2  + \frac{1}{2 v_d}\frac{\partial \Delta V}{\partial v_d} + 2\mu^2 \Bigr|_{M_{\rm IR}}. \label{eq:ewsb}
\end{eqnarray}
The soft masses $m_{H_u}^2$ and $m_{H_d}^2$ as well as the one-loop corrections to the Higgs potential $\Delta V$ are evaluated at the scale $M_{\rm IR}=\sqrt{m_{Q_3} m_{ {\bar U}_3} }$ (the stop mass scale).
We assume that the ratio between $m_{3/2}$ and $M_{1/2}$, $r = m_{3/2} / M_{1/2}$, is fixed by some high energy physics. 
Then, the EWSB scale $v$ is determined by three fundamental parameters,
$\mu|_{M_{\rm inp}}$, $M_{1/2}$ and $B_0$.

Now, we estimate the fine-tuning of the EWSB scale with respect to the fundamental parameters. 
We employ the following fine-tuning measure~\cite{ft_measure}:
\begin{eqnarray}
\Delta = \max_a\{ |\Delta_a| \}, \ \Delta_a = \Bigl\{ \frac{\partial  \ln v}{\partial \ln \mu} \Bigr|_{v=v_{\rm obs}}, \frac{\partial  \ln v}{\partial \ln M_{1/2}} \Bigr|_{v=v_{\rm obs}}, \frac{\partial  \ln v}{\partial \ln B_0} \Bigr|_{v=v_{\rm obs}} \Bigr\}, 
\end{eqnarray}
where $v_{\rm obs} \simeq 174.1$ GeV.

\subsection{The case for vanishing A-terms}

When A-terms vanish, the soft mass of the up-type Higgs at the IR scale can be written in terms of  $M_{1/2}$ and $m_{3/2}$. By numerically solving 2-loop renormalization group equations~\cite{Martin:1993zk},
it is given by
\begin{eqnarray}
m_{H_u}^2 (M_{\rm IR}=2{\rm \, TeV}) \simeq 0.689 m_{3/2}^2 - 1.182  M_{1/2}^2,
\end{eqnarray}
for $M_{\rm IR}=2$ TeV, 
and
\begin{eqnarray}
m_{H_u}^2 (M_{\rm IR}=3{\rm \, TeV}) \simeq 0.694 m_{3/2}^2 - 1.067  M_{1/2}^2,
\end{eqnarray}
for $M_{\rm IR}=3$ TeV. Here, the top pole mass
$m_t=173.34$ GeV, $\alpha_S(M_Z)=0.1184$ and $\tan\beta$=25.
We see that if  $r = m_{3/2}/M_{1/2} \simeq 6/5\mathchar`-4/3$, $m_{H_u}(M_{\rm IR})$ becomes significantly smaller than $m_{3/2}$ and $M_{1/2}$: the fine-tuning of the EWSB scale becomes mild. 
Since the contribution of the soft mass of the down-type Higgs to the EWSB scale is suppressed by $1/\tan^2 \beta$ (see Eq.\,(\ref{eq:ewsb})), it is less important than $m_{H_u}^2$ if $\tan\beta$ is large. 

\begin{figure}[t]
\begin{center}
\includegraphics[scale=1.00]{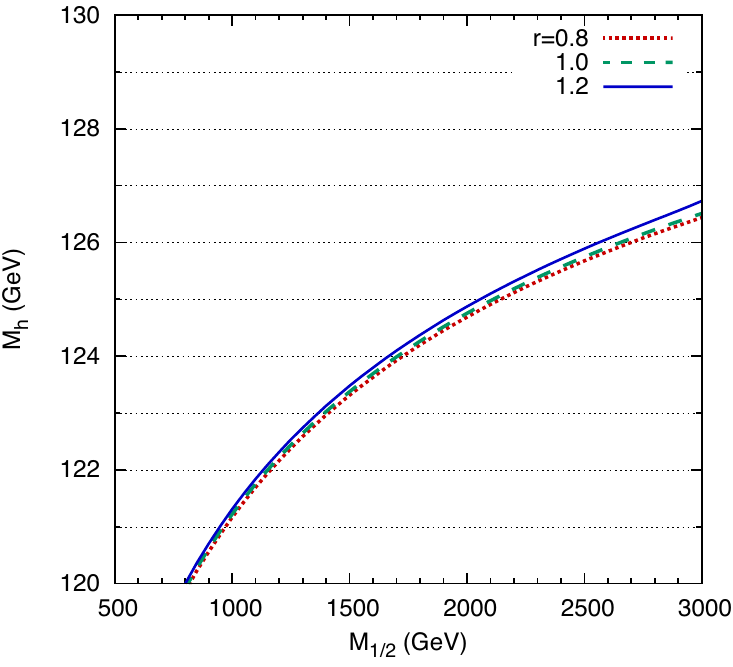}
\caption{The Higgs boson mass as a function of $M_{1/2}$. The ratio $r$ is defined by $r=m_{3/2}/M_{1/2}$.
We take $A_0=0$, $\tan\beta=25$, $\alpha_s(M_Z)=0.1184$ and $m_t=173.34$ GeV. 
}
\label{fig:fig_mh0}
\end{center}
\end{figure}

Let us estimate the required size of $M_{1/2}$ for the observed Higgs boson mass around 125 GeV.
In Fig.~\ref{fig:fig_mh0}, the Higgs boson mass is shown as a function of $M_{1/2}$.  
The Higgs boson mass is evaluated using {\tt FeynHiggs 2.10.3}~\cite{feynhiggs}. The mass spectrum of SUSY particles are calculated using {\tt Softsusy 3.5.2}~\cite{softsusy}.  
The Higgs boson mass $m_{h}=(123, 124, 125)$ GeV is obtained for $M_{1/2} \simeq (1400, 1700, 2100)$ GeV. Therefore, $M_{1/2}=1400\mathchar`-2100$ GeV is consistent with the observed Higgs boson mass. 
Note that $m_h$ is almost insensitive to $r=m_{3/2}/M_{1/2}$.

\begin{figure}[t]
\begin{center}
\includegraphics[scale=0.95]{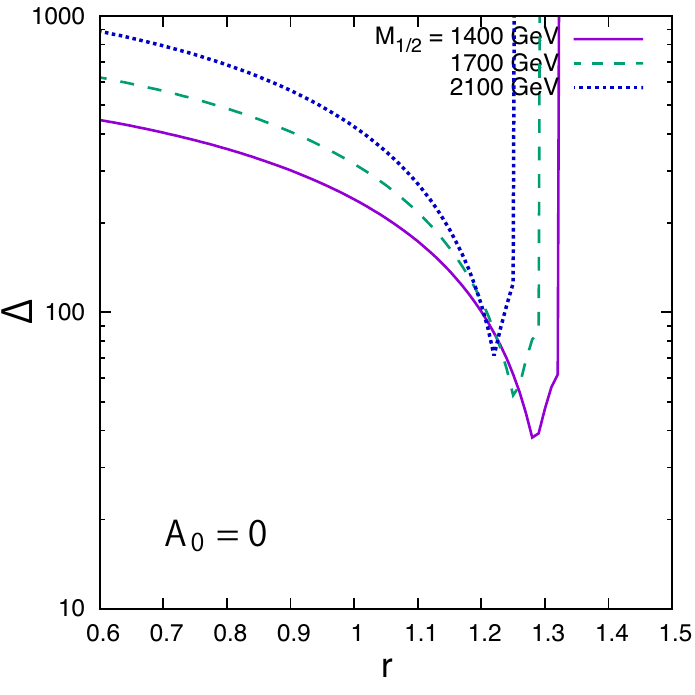}
\includegraphics[scale=0.95]{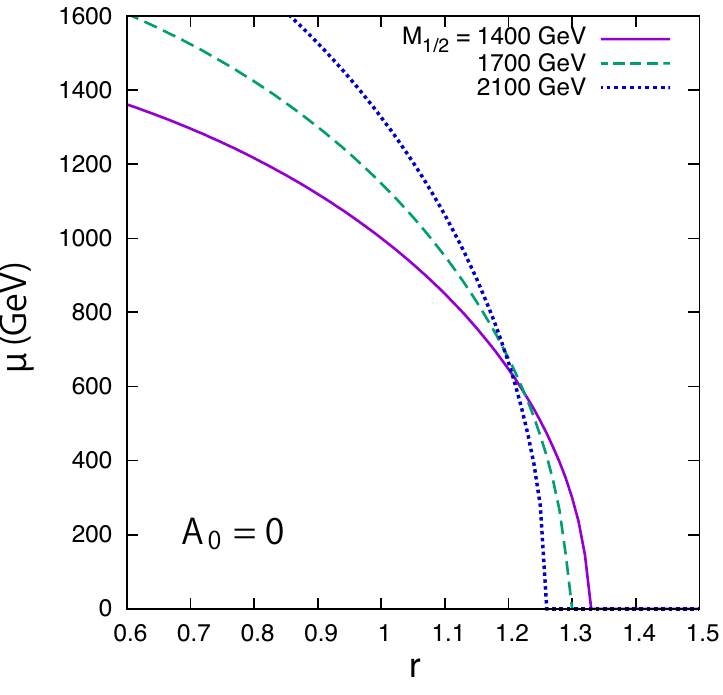}
\caption{$\Delta$ and $\mu$ as a function of $r$. In each panel, different curves correspond to different $M_{1/2}$. 
The other parameters are same as in Fig.~\ref{fig:fig_mh0}.
}
\label{fig:zero_a}
\end{center}
\end{figure}

In Fig.~\ref{fig:zero_a}, $\Delta$ and $\mu$ are shown.   
It is very much encouraging to see that there is indeed a parameter region where only a mild fine tuning is required ($\Delta \simeq 30\mathchar`-50$). In such a region around $r\simeq 6/5$, $\mu$ is small: 
the Higgsino  is the lightest SUSY particle (LSP) if $\mu$ is sufficiently small. 
In this case, the Higgsino is a dark matter candidate. 
With non-thermal productions~\cite{non-thermal}, the abundance of this Higgsino-like neutralino can be consistent with 
the observed dark matter abundance.
The spin-independent neutralino-nucleon cross section is around $10^{-45}$ cm$^2$ (see Table 1), and it is consistent with the current limit from the LUX experiment~\cite{lux}.

Apart from this small $\mu$ region, the stau is the LSP. However, we can enlarge the region of the Higgsino LSP, by introducing the small scalar masses for the sleptons (and squarks).
These small scalar masses
can be generated at the one-loop level due to the explicite breaking of $E_7/SU(5)\times U(1)^3$~\cite{Evans:2013uza,Gaillard:2000fk}, and the stau mass is lifted. 

The minimum value of $\Delta$ is found to be $\Delta=40\mathchar`-70$ for $M_{1/2}=1400\mathchar`-2100$ GeV.
The mild fine-tuning of $\Delta=40\mathchar`-70$ is consistent with the observed Higgs boson mass around 125 GeV,
as explained above.

\subsection{The case for non-vanishing A-terms}
When the VEV of $Z$ is of the order of the Planck mass scale, we have non-vanishing A-terms.\footnote{
The scalar potential contains the following term:
\begin{eqnarray}
V \ni e^{K/2} \frac{\partial K}{\partial Z} ( \lambda_{i j k} Q_i Q_j Q_k + \lambda_{l m} Q_l Q_m) \left(\frac{\partial W}{\partial Z} + \frac{\partial K}{\partial Z} W\right)^* + {\rm h.c.}\, ,
\end{eqnarray}
where $Q_i$ denotes a scalar component of a MSSM superfield.
Therefore, if $(\partial K/\partial Z)=Z \sim 1$,  the above term gives $O(m_{3/2})$ contribution to trilinear A-terms and bilinear B-terms.
}
In this case, $m_{H_u}^2$ at $M_{\rm IR}$ is written as
\begin{eqnarray}
m_{H_u}^2 (M_{\rm IR}=2{\rm \, TeV}) \simeq 0.689 m_{3/2}^2 - 1.182 M_{1/2}^2 + 0.331\, M_{1/2} A_0 - 0.120 A_0^2, \label{eq:non_zero_a1}
\end{eqnarray}
for $M_{\rm IR}=2$ TeV, and 
\begin{eqnarray}
m_{H_u}^2 (M_{\rm IR}=3{\rm \, TeV}) \simeq 0.694 m_{3/2}^2 - 1.067 M_{1/2}^2 + 0.322\, M_{1/2} A_0 - 0.109 A_0^2,
\label{eq:non_zero_a2}
\end{eqnarray}
for $M_{\rm IR}=3$ TeV, 
where $A_0$ is the universal trilinear couplings given at $M_{\rm inp}$. 
Note that the coefficients of $A_0^2$ and $M_{1/2} A_0$ are not large.
Therefore, as long as $|A_0| < M_{1/2}$, the presence of $A_0$ does not affect the fine-tuning of the EWSB scale significantly. 

Since we have non-zero A-terms, the fine-tuning measure $\Delta$ becomes
\begin{eqnarray}
\Delta = \max_a\{ |\Delta_a| \}, \ \Delta_a = \Bigl\{ \frac{\partial  \ln v}{\partial \ln \mu}, \frac{\partial  \ln v}{\partial \ln M_{1/2}}, \frac{\partial  \ln v}{\partial \ln A_0}, \frac{\partial  \ln v}{\partial \ln B_0} \Bigr\} .
\end{eqnarray}

\begin{figure}[t]
\begin{center}
\includegraphics[scale=1.00]{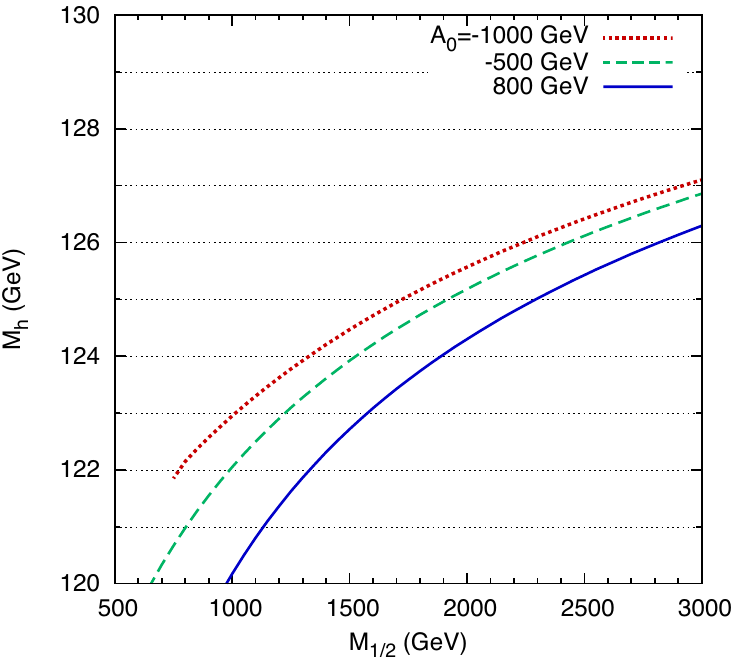}
\caption{The Higgs boson mass as a function of $M_{1/2}$ for different $A_0$. Here, $r=1.1$ and the other parameters are same as in Fig.~\ref{fig:fig_mh0}.
}
\label{fig:fig_mh1}
\end{center}
\end{figure}

In Fig.~\ref{fig:fig_mh1},
the Higgs boson mass is shown with $A_0 \neq 0$.
When $A_0$ is negative (positive), required $M_{1/2}$ for the Higgs boson mass becomes smaller (larger). The Higgs boson mass of  $m_h=(123,124,125)$ GeV is obtained for $M_{1/2} \simeq (1200, 1500, 1900)$ GeV and $A_0=-500$ GeV, while $m_h=(123,124,125)$ GeV is obtained for $M_{1/2} \simeq (1600, 1900, 2300)$ GeV and $A_0=800$ GeV.

\begin{figure}[t]
\begin{center}
\includegraphics[scale=0.95]{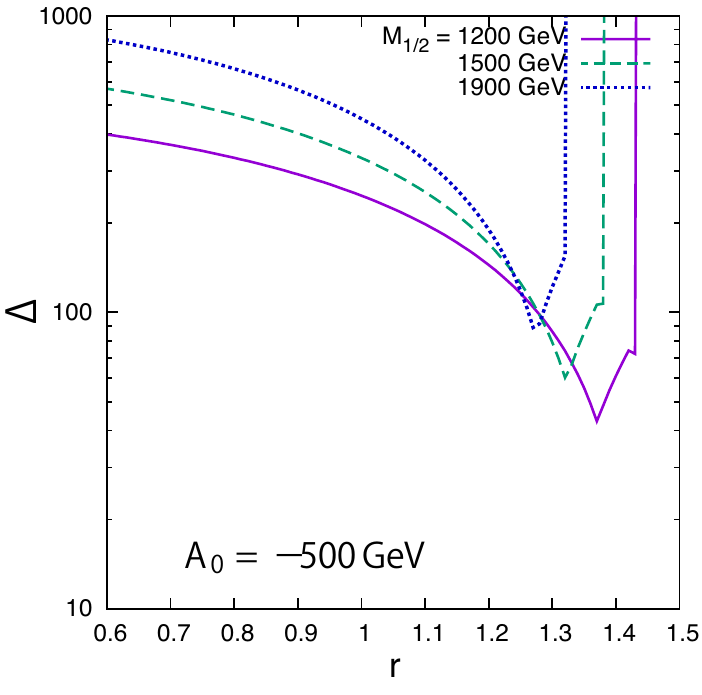}
\includegraphics[scale=0.95]{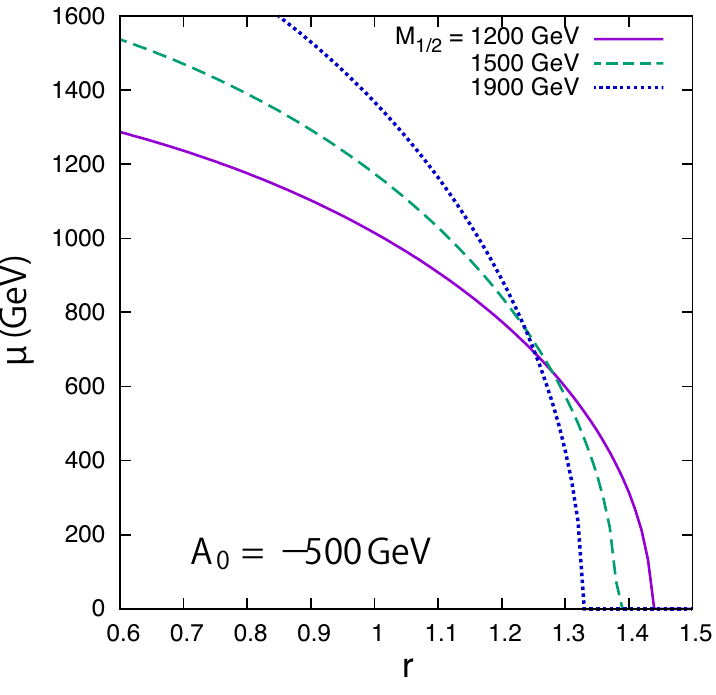}
\includegraphics[scale=0.95]{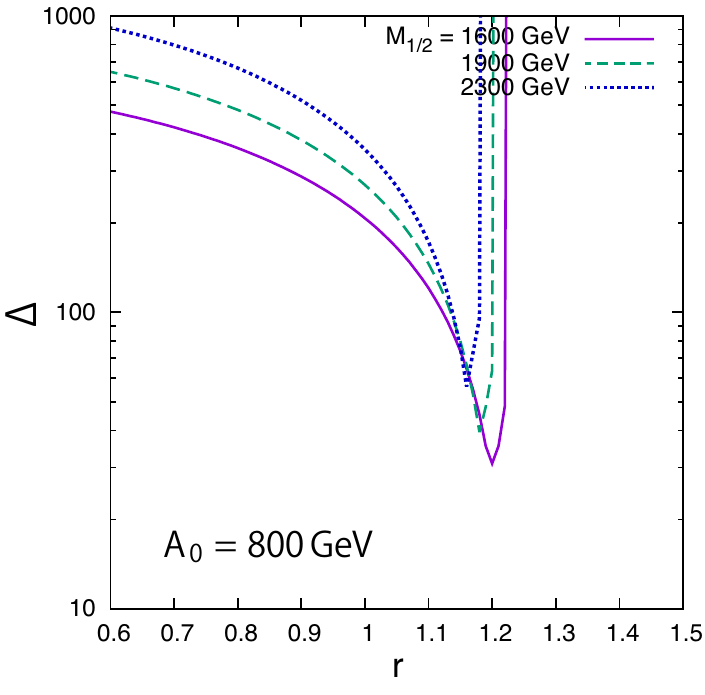}
\includegraphics[scale=0.95]{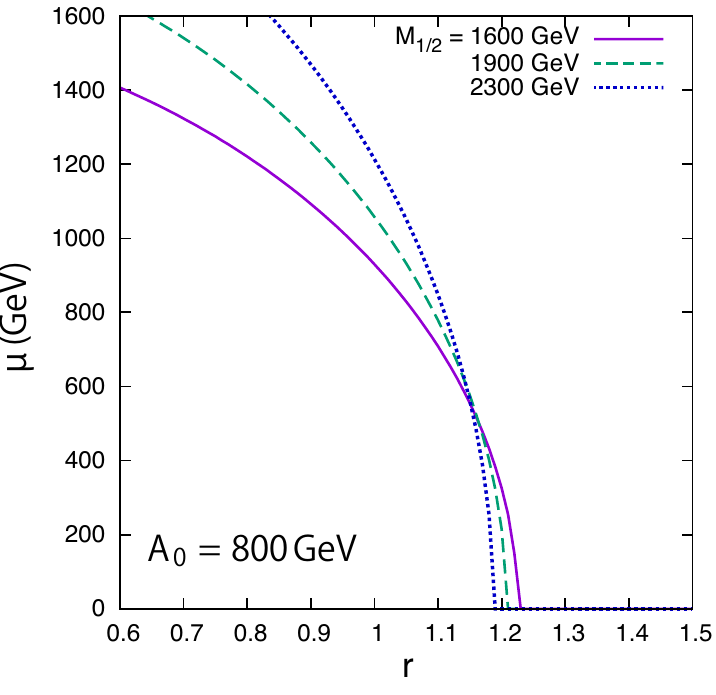}
\caption{$\Delta$ and $\mu$ as a function of $r$ for $A_0 \neq 0$. In the upper (lower) two panels, $A_0=-500$ (800) GeV. The other parameters are same as in Fig.~\ref{fig:fig_mh0}.
}
\label{fig:non_zero_a}
\end{center}
\end{figure}

In Fig.~\ref{fig:non_zero_a}, $\Delta$ and $\mu$ are shown for non-zero A-terms. In the upper (lower) two panels, $A_0=-500$ (800) GeV. For $A_0 = -500$ GeV, $\Delta=40\mathchar`-90$: although smaller $M_{1/2}$ is allowed, the fine-tuning becomes slightly worse than that of the model with $A_0=0$.
On the other hand, for $A_0 = 800$ GeV,  $\Delta=30\mathchar`-60$. The positive $A_0$ slightly reduces $\Delta$ (see Eq.~(\ref{eq:non_zero_a1})(\ref{eq:non_zero_a2})) compared to the model with $A_0=0$. The larger $A_0$ is not favored, since $(\partial \ln v/\partial \ln A_0)$ becomes large and so does $\Delta$.

\begin{figure}[t]
\begin{center}
\includegraphics[scale=1.00]{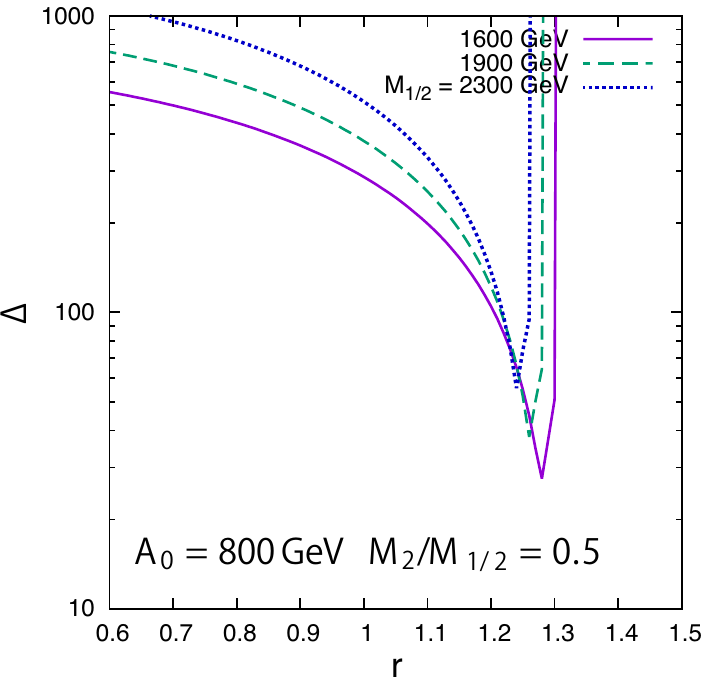}
\includegraphics[scale=1.00]{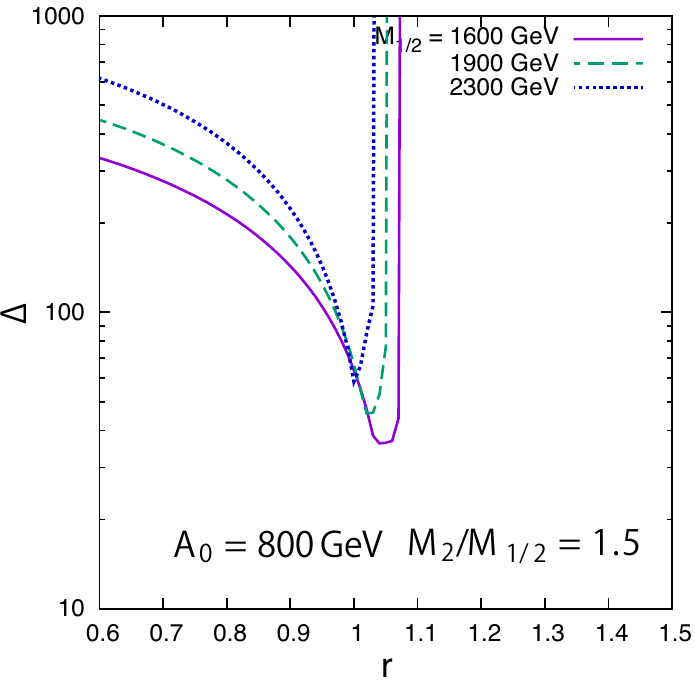}
\caption{$\Delta$ in the non-universal gaugino mass cases. In the left (right) panel $M_2/M_{1/2}$=0.5 (1.5). 
Here, $A_0=800$ GeV.
The other parameters are same as in Fig.~\ref{fig:fig_mh0}.
}
\label{fig:delta_non_univ}
\end{center}
\end{figure}

\vspace{15pt}
We note that the focus point of the EWSB scale is maintained, even
if the gaugino masses are non-universal. In Fig.~\ref{fig:delta_non_univ}, $\Delta$ is shown for $M_2/M_{1/2}=0.5$ and $M_2/M_{1/2}=1.5$, where $M_2$ is the wino mass at $M_{\rm inp}$. The gluino and bino masses are taken as $M_{1/2}$. The ratio $M_{2}/M_{1/2}$ is assumed to be fixed at $M_{\rm inp}$. Although $r=m_{3/2}/M_{1/2}$ giving the minimum value of $\Delta$ is slightly shifted from that of the universal gaugino mass case, it can be seen that the small $\Delta$ is still maintained even if the gaugino masses are non-universal.

\begin{figure}[t]
\begin{center}
\includegraphics[scale=1.00]{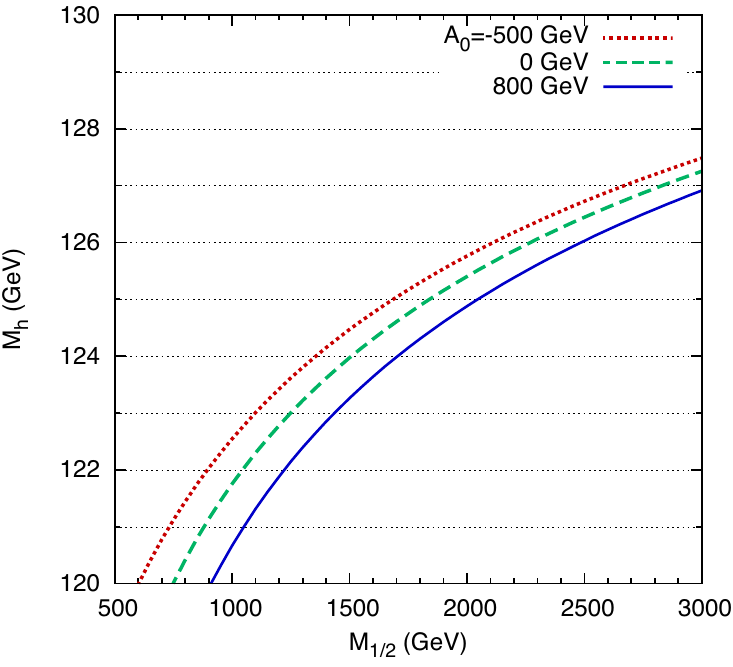}
\caption{The Higgs boson mass for the larger top mass, $m_t=174.10$ GeV. Here, $r=1.1$, $\tan\beta=25$.
}
\label{fig:mh_174}
\end{center}
\end{figure}

\begin{figure}[t]
\begin{center}
\includegraphics[scale=1.00]{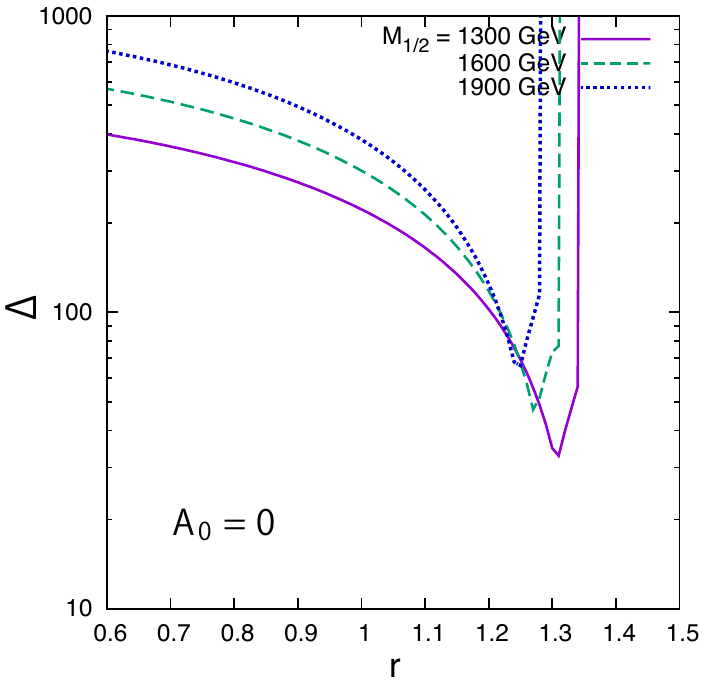}
\includegraphics[scale=1.00]{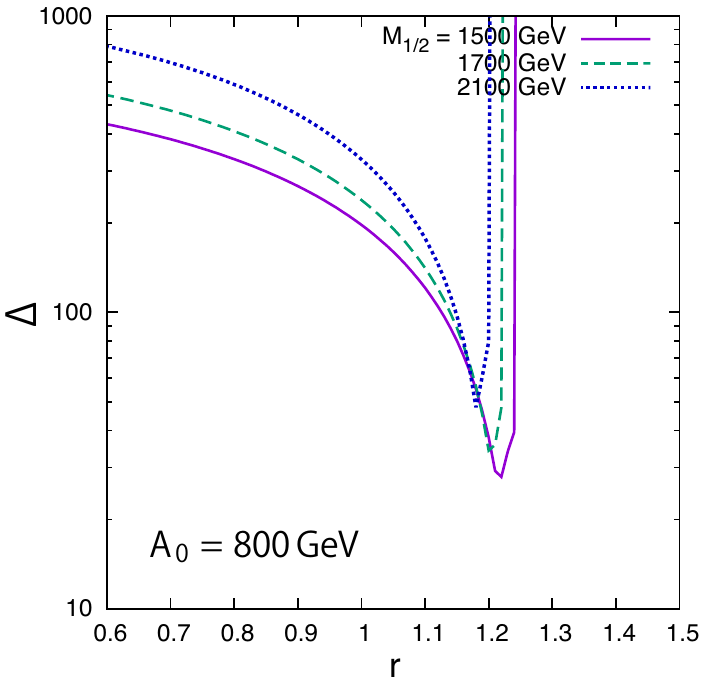}
\caption{$\Delta$ for the larger top mass, $m_t=174.10$ GeV. In the right (left) panel, $A_0=800$ (0) GeV. The other parameters are same in Fig.~\ref{fig:mh_174}.
}
\label{fig:174_delta}
\end{center}
\end{figure}

Fig.~\ref{fig:mh_174} and \ref{fig:174_delta} show
$m_h$ and $\Delta$ for the larger top mass, $m_t=174.10$ GeV.
We see that the fine-tuning is slightly improved as $\Delta=30\mathchar`-50$ for $A_0=800$ GeV and $M_{1/2}=1500\mathchar`-2100$ GeV,
where the Higgs mass of $m_h=123\mathchar`-125$ GeV is obtained.

\begin{figure}[t]
\begin{center}
\includegraphics[scale=1.00]{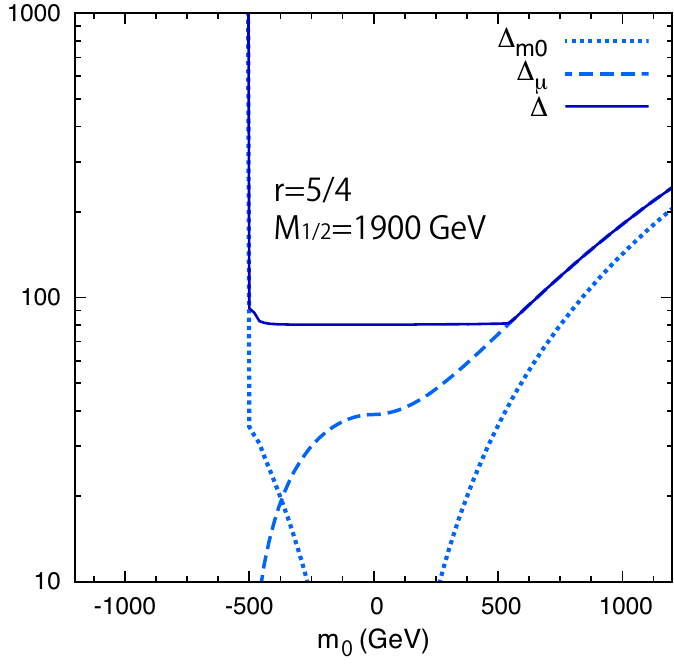}
\caption{$\Delta$ as a function of $m_0$. Squark masses are $m_{Q}^2=m_{\bar{U}}^2=m_{\bar D}^2 = {\rm sign} (m_{0}) |m_{0}|^2$, and slepton masses are $m_{L}^2=m_{\bar E}^2=|m_0|^2$. The other parameters are $r=5/4$, $M_{1/2}=1900$ GeV, $\tan\beta=25$, $A_0=0$ and $m_t=173.34$ GeV.
}
\label{fig:m0}
\end{center}
\end{figure}

Finally, we discuss the stability of our focus point against small changes of sfermions masses, since
one-loop threshold corrections may generate sfermions masses of $O(100)$ GeV~\cite{Gaillard:2000fk,Evans:2013uza}.
In Fig.~\ref{fig:m0}, we show $\Delta$ when non-zero squark and slepton masses $m_0^2$ exist.
Here, we have also considered the contribution of $m_0^2$ to fine-tuning,
$\Delta_{m_0}=\partial \ln v/\partial \ln |m_0|$.
We see that the focus point is maintained,
as long as $m_0 \lsim 500$ GeV.

\begin{table}[t!]
  \begin{center}
  \small
    \begin{tabular}{  c | c  }
            {\bf P1} & \\
\hline
    $M_{1/2} $ & 1400 GeV \\
    $r $ & 1.30  \\
    $A_0$ & 0 GeV\\
    $\tan \beta$ & 25 \\
    \hline
\hline
    $\mu$ & 300 \\
    $\Delta$ & 47 \\    
    $m_{\rm gluino}$ & 2.97 TeV \\
      $m_{\tilde{q}}$ & 2.56\,-\,2.69 TeV \\
    $m_{\tilde{t}_{1,2}}$ & 1.94, 2.38 TeV \\
    $m_{\tilde{\tau}_1}$ & 322 GeV\\
     $m_{\chi_1^0}$, $m_{\chi_2^0}$ & 299, 310 GeV \\
     $m_{\chi_3^0}$, $m_{\chi_4^0}$ & 611, 1142 GeV \\
     $m_{\chi_1^{\pm}}$, $m_{\chi_2^{\pm}}$ & 305, 1142 GeV \\
     $m_{h}$ & 123.2 GeV \\
     $(\sigma_p)_{\rm SI}$ & $2.5 \cdot 10^{-45}$ cm$^2$ \\
     & 
    \end{tabular}
        \hspace{20pt}
    \begin{tabular}{  c | c  }
            {\bf P2} & \\
\hline
    $M_{1/2} $ & 1900 GeV \\
    $r $ & 1.24  \\
    $A_0$ & 0 GeV\\
    $\tan \beta$ & 25 \\
    \hline
\hline
    $\mu$ & 471 \\
    $\Delta$ & 66 \\    
    $m_{\rm gluino}$ & 3.95 TeV \\
      $m_{\tilde{q}}$ & 3.39\,-\,3.57 TeV \\
    $m_{\tilde{t}_{1,2}}$ & 2.60, 3.16 TeV \\
    $m_{\tilde{\tau}_1}$ & 457 GeV\\
     $m_{\chi_1^0}$, $m_{\chi_2^0}$ & 475, 484 GeV \\
     $m_{\chi_3^0}$, $m_{\chi_4^0}$ & 834, 1553 GeV \\
     $m_{\chi_1^{\pm}}$, $m_{\chi_2^{\pm}}$ & 480, 1554 GeV \\
     $m_{h}$ & 124.7 GeV \\
     $(\sigma_p)_{\rm SI}$ & $-$ \\
     & 
    \end{tabular}
    \begin{tabular}{  c | c  }
            {\bf P3} & \\
\hline
    $M_{1/2} $ & 2300 GeV \\
    $r $ & 1.16 \\
    $A_0$ & 800 GeV\\
    $\tan \beta$ & 25 \\
    \hline
\hline
    $\mu$ & 483 \\
    $\Delta$ & 56 \\    
    $m_{\rm gluino}$ & 4.73 TeV \\
      $m_{\tilde{q}}$ & 4.04\,-\,4.26 TeV \\
    $m_{\tilde{t}_{1,2}}$ & 3.19, 3.81 TeV \\
    $m_{\tilde{\tau}_1}$ & 602 GeV\\
     $m_{\chi_1^0}$, $m_{\chi_2^0}$ & 491, 497 GeV \\
     $m_{\chi_3^0}$, $m_{\chi_4^0}$ & 1013, 1882 GeV \\
     $m_{\chi_1^{\pm}}$, $m_{\chi_2^{\pm}}$ & 494, 1882 GeV \\
     $m_{h}$ & 125.1 GeV \\
    $(\sigma_p)_{\rm SI}$ & $0.8 \cdot 10^{-45}$ cm$^2$ \\
    \end{tabular}
        \hspace{15pt}
    \begin{tabular}{  c | c  }
            {\bf P4} & \\
\hline
    $M_{1/2} $ & 1600 GeV \\
    $r $ & 1.20 \\
    $A_0$ & 800 GeV\\
    $\tan \beta$ & 25 \\
    \hline
\hline
    $\mu$ & 326 \\
    $\Delta$ & 31 \\    
    $m_{\rm gluino}$ & 3.28  TeV \\
      $m_{\tilde{q}}$ & 2.89\,-\,3.04 TeV \\
    $m_{\tilde{t}_{1,2}}$ & 2.28, 2.73 TeV \\
    $m_{\tilde{\tau}_1}$ & 408 GeV\\
     $m_{\chi_1^0}$, $m_{\chi_2^0}$ & 328, 337 GeV \\
     $m_{\chi_3^0}$, $m_{\chi_4^0}$ & 699, 1305 GeV \\
     $m_{\chi_1^{\pm}}$, $m_{\chi_2^{\pm}}$ & 333, 1305 GeV \\
     $m_{h}$ & 123.2 GeV \\
     $(\sigma_p)_{\rm SI}$ & $1.8 \cdot 10^{-45}$ cm$^2$ \\
    \end{tabular}
   \caption{The SUSY mass spectra. Here, $m_t=173.34$ GeV. 
   The spin-independent neutralino-proton cross section, $(\sigma_p)_{\rm SI}$, is calculated using {\tt micrOMEGAs 4.1.7}~\cite{omega}, with $f_s \simeq 0.045$ ~\cite{Junnarkar:2013ac}.
   }
  \label{table:spectrum1}
  \end{center}
\end{table}

\section{Discussion and conclusions}

We have shown the presence of a new focus point based on the $E_7/SU(5) \times U(1)^3$ NLS model. 
With the fixed ratio of the gravitino mass to the gaugino mass around $5/4$, the EWSB scale is insensitive to the soft SUSY breaking mass scale. Since all the soft masses apart from those of the Higgs doublets are mainly generated from gaugino loops, this focus point scenario is free from the SUSY flavor problem. 
Small fine-tuning, $\Delta=30\mathchar`-70$, is consistent with the observed Higgs boson mass around 125 GeV.
On the focus point,
the gluino and squark masses are predicted around 3-4 TeV, as shown in Table~\ref{table:spectrum1}.
Since squarks lighter than 3.5 TeV (3.0 TeV) can be excluded (discovered) with the 3000 fb$^{-1}$ data for the gluino mass of 4.5 TeV at the LHC~\cite{lhc_high_lumi}, it is expected that the present scenario can be tested at the high luminosity LHC.

The Higgsino-like neutralino is the LSP in the region with mild fine-tuning (i.e.~small $\Delta$).
This neutralino can be dark matter: the observed dark matter abundance may be explained by some non-thermal dark matter production.
The spin-independent neutralino-nucleon cross section is around $10^{-45}$ cm$^2$; therefore, the neturalino dark matter can be easily discovered/excluded at future direct detection experiments.

Let us comment on focus points in general.
The EWSB scale is basically determined by
$m_{H_{u,d}}^2$,  $m_{\rm sfermion}^2$ and $M_{\rm gaugino}$.
Focus points, or, seminatual SUSY, are based on postulated relations between those parameters.
The focus point discussed in~\cite{Feng:1999mn} assumes the universal scalar masses and small gaugino masses (see also~\cite{Brummer:2013dya}).
The focus point in~\cite{Yanagida:2013ah} assumes vanishing scalar masses and a specific ratio between the wino and the gluino masses.
The focus point in this paper assumes vanishing sfermion masses motivated from the NLS model.%
\footnote{
Thus, our focus point also exists in the gaugino mediation model of~\cite{Kaplan:1999ac,Chacko:1999mi},
where sfermion masses vanish.
}
We have found the presence of a focus point when the ratio of $m_{H_u}^2$ to the gluino mass is fixed around $5/4$.
We do not have concrete high energy physics models which lead to these relations at present.
However, taking the naturalness problem seriously,
it would be helpful to search for focus points phenomenologically and examine their predictions.
Once the predictions are confirmed by experiments,
we hope that the nature of the focus points will guide us to unknown high energy physics.

Finally, let us comment on cosmological aspects of our model. In our model, the gravitino is as heavy as a few TeV, and  it decays into standard model particles and their superpartners with a long life-time; therefore, we need to pay attention to the cosmological gravitino problem~\cite{Weinberg:1982zq}. However, in fact, 
the gravitino problem is avoided if the reheating temperature is lower than about $10^6$ GeV~\cite{Kawasaki:2008qe}.

We have two modulus fields, the SUSY breaking field $Z$ and the chiral multiplet $S$.
They may obtain large amplitudes and hence large energy densities in the early universe.
Decay of moduli ruins the success of the Big Bang Nucleosynthesis (BBN) and produces large entropy~\cite{Coughlan:1983ci} as well as too much LSP dark matter.
However, the amplitude of $Z$ can be suppressed by couplings of $Z$ in the superpotential~\cite{Harigaya:2013ns}
or by strong couplings with the inflaton in the Kahler potential~\cite{Linde:1996cx,Takahashi:2010uw,Nakayama:2011wqa}.
The latter solution is also applicable to $S$.\footnote{
The former one is not applicable to $S$ in this framework;
if the superpotential depends on $S$, soft masses of squarks and sleptons no longer vanish (see Sec.~\ref{sec:soft mass}).}

The imaginary component of the chiral multiplet $S$ does not obtain its mass from the Kahler potential due to the shift symmetry of $S$ (see Eq.~(\ref{eq:E7 Kahler})), which is a U(1) part of the $E_7$ symmetry~\cite{Kugo:2010fs}.
If the imaginary component remains massless and is produced in the early universe,  it may contribute to the dark radiation of the universe.
It is also possible that the shift symmetry is anomalous and hence obtains its mass from QCD dynamics~\cite{Iwamoto:2014ywa}.
Then the imaginary component works as a QCD axion~\cite{Peccei:1977hh} and hence solves the strong CP problem.

In the NLS model, not only soft mass squared but also Hubble induced masses vanish.
Then squarks and sleptons obtain unsuppressed quantum fluctuations during inflation.
It would be interesting to investigate dynamics of squarks and sleptons in the early universe.

In the above discussion on cosmology, we have assumed that the gravitino mass is $O(1)$ TeV.
It would be interesting to consider a model with a gravitino mass far larger than $O(1)$ TeV, say $O(100)$ TeV.
In this case, the moduli and the gravitino decay well before the BBN and hence is free from the constraint from the BBN.%
\footnote{
The entropy production is not a problem if baryogenesis is very efficient, as is the case of Affleck-Dine baryogenesis~\cite{Affleck:1984fy}.
The production of the LSP is also not a problem if the R-parity is violated.
Even if the R-parity is not violated, small amplitudes of moduli may be explained by the anthropic principle.
}
If the SUSY breaking field $Z$ weakly couples to the Higgs fields $H_u$ and $H_d$
in the conformal frame of the Kahler potential and to gauge multiples in gauge kinetic functions, we obtain a similar focus point as what we have discussed in this paper.%
\footnote{
With a gravitino mass of $O(100)$ TeV, the anomaly mediation in general generates soft masses of $O(1)$ TeV~\cite{Randall:1998uk} (see also~\cite{Bagger:1999rd}).
Our focus point seems to be ruined by the anomaly mediated soft masses.
In NLS models, however, the anomaly mediation can be suppressed~\cite{HIYY}.
}

\section*{Acknowledgments}
This work is supported by Grant-in-Aid for Scientific research from the
Ministry of Education, Science, Sports, and Culture (MEXT), Japan,
No.\ 26104009 and 26287039 (T.\,T.\,Y.),
and also by World Premier International Research Center Initiative (WPI Initiative), MEXT, Japan (K.\,H.~and T.\,T.\,Y.).
The research leading to these results has received funding
from the European Research Council under the European Unions Seventh
Framework Programme (FP/2007-2013) / ERC Grant Agreement n. 279972
``NPFlavour'' (N.\,Y.).
The work of K.\,H.\, is supported in part by a JSPS Research Fellowships for Young Scientists.

\end{document}